\documentclass[a4paper]{jpconf}
\usepackage{graphicx}

\newcommand{\be}{\begin{equation}}
\newcommand{\ee}{\end{equation}}
\newcommand{\bea}{\begin{eqnarray}}
\newcommand{\eea}{\end{eqnarray}}

 \def\bean{\begin{eqnarray*}}
 \def\eean{\end{eqnarray*}}

 \def\bm#1{\mbox{\boldmath$#1$}}
 \def\gsim{\mathrel{\rlap{\lower0.2em\hbox{$\sim$}}\raise0.2em\hbox{$>$}}}
 \def\ksim{\mathrel{\rlap{\lower0.2em\hbox{$\sim$}}\raise0.2em\hbox{$<$}}}

\begin{document}
\title{Covariant transport approach for strongly interacting partonic systems}

\author{W. Cassing$^1$ and E. L. Bratkovskaya$^2$}

\address{$^1$Institut f\"ur Theoretische Physik, %
  Universit\"at Giessen, %
  Germany}
\address{$^2$Institut f\"ur Theoretische Physik, %
  Universit\"at Frankfurt am Main, %
  Germany}

\ead{Wolfgang.Cassing@theo.physik.uni-giessen.de}

\begin{abstract}
The dynamics of partons, hadrons and strings in relativistic
nucleus-nucleus collisions is analyzed within the novel Parton-Hadron-String
Dynamics (PHSD) transport
approach, which is based on a dynamical quasiparticle model for
partons (DQPM) matched to reproduce recent lattice-QCD results - including
the partonic equation of state - in thermodynamic equilibrium.
Scalar- and vector-interaction densities are extracted from the
DQPM as well as effective scalar- and vector-mean fields for the
partons.  The transition from partonic to hadronic degrees of
freedom is described by covariant transition rates for the fusion
of quark-antiquark pairs or three quarks (antiquarks),
respectively, obeying flavor current-conservation, color
neutrality as well as energy-momentum conservation. Since the
dynamical quarks and antiquarks become very massive close to the
phase transition, the formed resonant 'pre-hadronic' color-dipole
states ($q\bar{q}$ or $qqq$) are of high invariant mass, too, and
sequentially decay to the groundstate meson and baryon octets
increasing the total entropy. When applying the PHSD approach to Pb+Pb colllisions
at 158 A$\cdot$GeV we find a significant effect of the partonic phase on the
production of multi-strange antibaryons due to a slightly enhanced $s{\bar
s}$ pair production from massive time-like
gluon decay and a larger formation of antibaryons in the hadronization process.

\end{abstract}

\section{Introduction}
The 'Big Bang' scenario implies that in the first micro-seconds of
the universe the entire state has emerged from a partonic system
of quarks, antiquarks and gluons -- a quark-gluon plasma (QGP) --
to color neutral hadronic matter consisting of interacting
hadronic states (and resonances) in which the partonic degrees of
freedom are confined. The nature of confinement and the dynamics
of this phase transition has motivated a large community for
several decades  and is still an outstanding question of todays
physics. Early concepts of the QGP were guided by the idea of a
weakly interacting system of partons which might be described by
perturbative QCD (pQCD). However, experimental observations at the
Relativistic Heavy Ion Collider (RHIC) indicated that the new
medium created in ultrarelativistic Au+Au collisions is
interacting more strongly than hadronic matter and consequently this concept had to be
severely questioned. Moreover, in line with theoretical studies in
Refs. \cite{Shuryak,Thoma,Andre} the medium showed phenomena of an
almost perfect liquid of partons \cite{STARS,Miklos3} as extracted
from the strong radial expansion and the scaling of elliptic flow
$v_2(p_T)$ of mesons and baryons with the number of constituent quarks
and antiquarks \cite{STARS}.

The question about the  properties of this (nonperturbative) QGP
liquid is discussed controversially in the literature  and
dynamical concepts describing the formation of color neutral
hadrons from colored partons are scarce. A fundamental
issue for hadronization models is the conservation of 4-momentum
as well as the entropy problem because by fusion/coalescence of
massless (or low constituent mass) partons to color neutral bound
states of low invariant mass (e.g. pions) the number of degrees of
freedom and thus the total entropy is reduced in the hadronization
process \cite{Koal1,Koal2,AMPT}. This problem - a violation of the
second law of thermodynamics  as well as  the conservation of
four-momentum and flavor currents - has been addressed in Ref.
\cite{PRC08} on the basis of the DQPM employing covariant
transition rates for the fusion of 'massive' quarks and antiquarks
to color neutral hadronic resonances or strings. In fact, the
dynamical studies for an expanding partonic fireball in Ref.
\cite{PRC08} suggest that the latter problems have come to a practical
solution.

A consistent dynamical approach - valid also for strongly
interacting systems - can be formulated on the basis of
Kadanoff-Baym (KB) equations \cite{Sascha1} or off-shell
transport equations in phase-space representation, respectively
\cite{Sascha1,Juchem,Knoll1}. In the KB theory the field quanta
are described in terms of dressed propagators with complex selfenergies.
Whereas the real part of the selfenergies can be related to
mean-field potentials (of Lorentz scalar, vector or tensor type),
the imaginary parts  provide information about the lifetime and/or
reaction rates of time-like 'particles' \cite{Crev}. Once the
proper (complex) selfenergies of the degrees of freedom are known
the time evolution of the system is fully governed  by off-shell
transport equations (as described in Refs. \cite{Sascha1,Crev}).
The determination/extraction of complex selfenergies for the
partonic degrees of freedom has been performed before in Refs.
\cite{Cassing06,Cassing07} by fitting lattice QCD (lQCD)
'data' within  the Dynamical QuasiParticle Model (DQPM). In fact,
the DQPM allows for a simple and transparent interpretation of
lattice QCD results for thermodynamic quantities as well as
correlators and leads to effective strongly interacting partonic
quasiparticles with broad spectral functions.  For a review on
off-shell transport theory and results from the DQPM in comparison
to lQCD we refer the reader to Ref. \cite{Crev}.

In the present contribution we will report on an update of the DQPM parameters
\cite{CaBra09} since more precise lQCD calculations for 2+1 flavors with almost
physical quark masses have become available in 2008
\cite{Cheng08}. Furthermore, we extend the study in Refs.
\cite{Cassing06,Cassing07} by fixing independently scalar- and
vector-interaction densities for the fermion degrees of freedom.
First applications of the PHSD approach to Pb +Pb collisions at SPS energies
- in comparison to
experimental data from the NA49 Collaboration - are presented, too.

\section{The PHSD approach}
The Parton-Hadron-String-Dynamics (PHSD) approach is a microscopic
covariant transport model that incorporates effective partonic as
well as hadronic degrees of freedom and involves a dynamical
description of the hadronization process from partonic to hadronic
matter. Whereas the hadronic part is essentially equivalent to the
conventional Hadron-Strings-Dynamics (HSD) approach \cite{HSD} the
partonic dynamics is based on the Dynamical QuasiParticle Model
(DQPM) \cite{Cassing06,Cassing07} which describes QCD
properties in terms of single-particle
Green's functions (in the sense of a two-particle
irreducible (2PI) approach).

\subsection{Reminder of the DQPM}
We briefly recall the basic assumptions of the DQPM: Following
Ref. \cite{Andre05} the dynamical quasiparticle mass (for gluons
and quarks) is assumed to be given by the thermal mass in the
asymptotic high-momentum regime, which is proportional to the
coupling $g(T/T_c)$ and the temperature  $T$, i.e. for gluons
\begin{equation}
 M_g^2(T) = \frac{g^2(T/T_c)}{6} \left( (N_c + \frac{1}{2}N_f)\, T^2
 + \frac{N_c}{2} \sum_q \frac{\mu_q^2}{\pi^2}
 \right) \, ,
 \label{eq:M2} \end{equation}
and for quarks (assuming vanishing constituent masses here) as,
\begin{equation}
M_q^2(T) = \frac{N_c^2-1}{8 N_c}\, g^2(T/T_c) \left( T^2 +
\frac{\mu_q^2}{\pi^2} \right) \, ,\label{eq:M2b} \end{equation}
with a running coupling (squared) (for $T > T_s$),
\begin{equation}
 g^2(T/T_c) = \frac{48\pi^2}{(11N_c - 2 N_f)  \ln(\lambda^2(T/T_c-T_s/T_c)^2}\ ,
 \label{eq:g2}
\end{equation} which permits for an enhancement near the critical temperature $T_c$.
Here $N_c = 3$ stands for the number of colors while $N_f$ denotes
the number of flavors. The parameters controlling the infrared
enhancement of the coupling $\lambda $ and $T_s$
have been refitted compared to Ref. \cite{Andre05} to the recent
lQCD results from Ref. \cite{Cheng08} (see below).

 The width  for gluons and quarks (for $\mu_q = 0$) is adopted in
the form \cite{Pisar89LebedS}
\begin{equation}
  \gamma_g(T)
  =
  N_c \frac{g^2 T}{8 \pi} \,  \ln\frac{2c}{g^2} \, , \hspace{2cm}
    \gamma_q(T)
  =
  \frac{N_c^2-1}{2 N_c} \frac{g^2 T}{8 \pi} \,  \ln\frac{2c}{g^2}
  \,.
 \label{eq:gamma}
\end{equation} where the parameter $c$  is related to a
magnetic cut-off. We stress that a non-vanishing width $\gamma$ is the central difference
of the DQPM to conventional quasiparticle models
\cite{qp1,qp2}. Its influence is essentially seen in
correlation functions as e.g. in the stationary limit of the
correlation function in the off-diagonal elements of the
energy-momentum tensor $T^{kl}$ which defines the shear viscosity
$\eta$ of the medium \cite{Andre,Kubo}. Here a sizable width is
mandatory to obtain a small ratio in the shear viscosity to
entropy density $\eta/s$ \cite{Andre}.

\begin{figure}[t]
\includegraphics*[width=72mm]{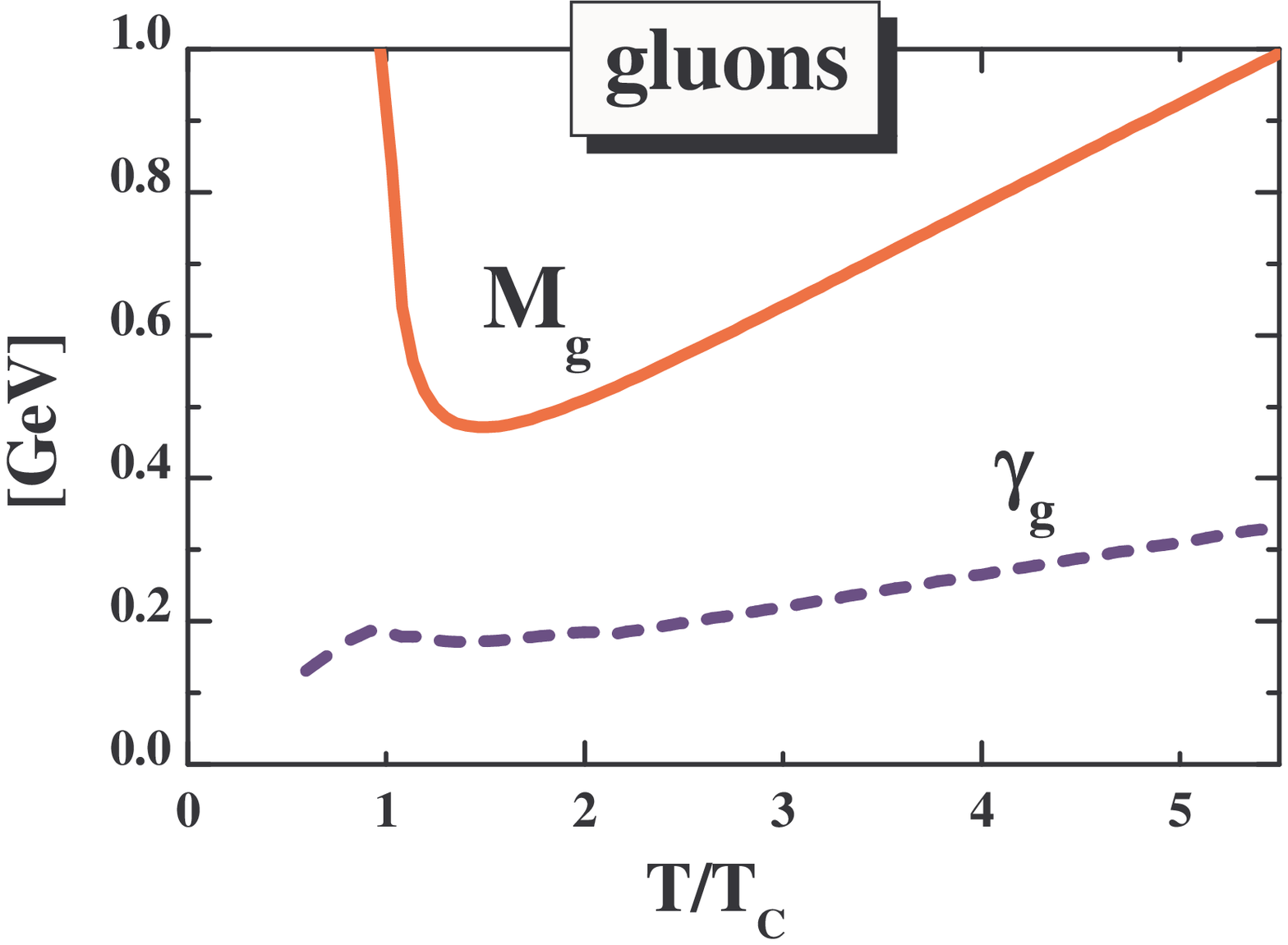}\hspace*{3mm}
\includegraphics*[width=72mm]{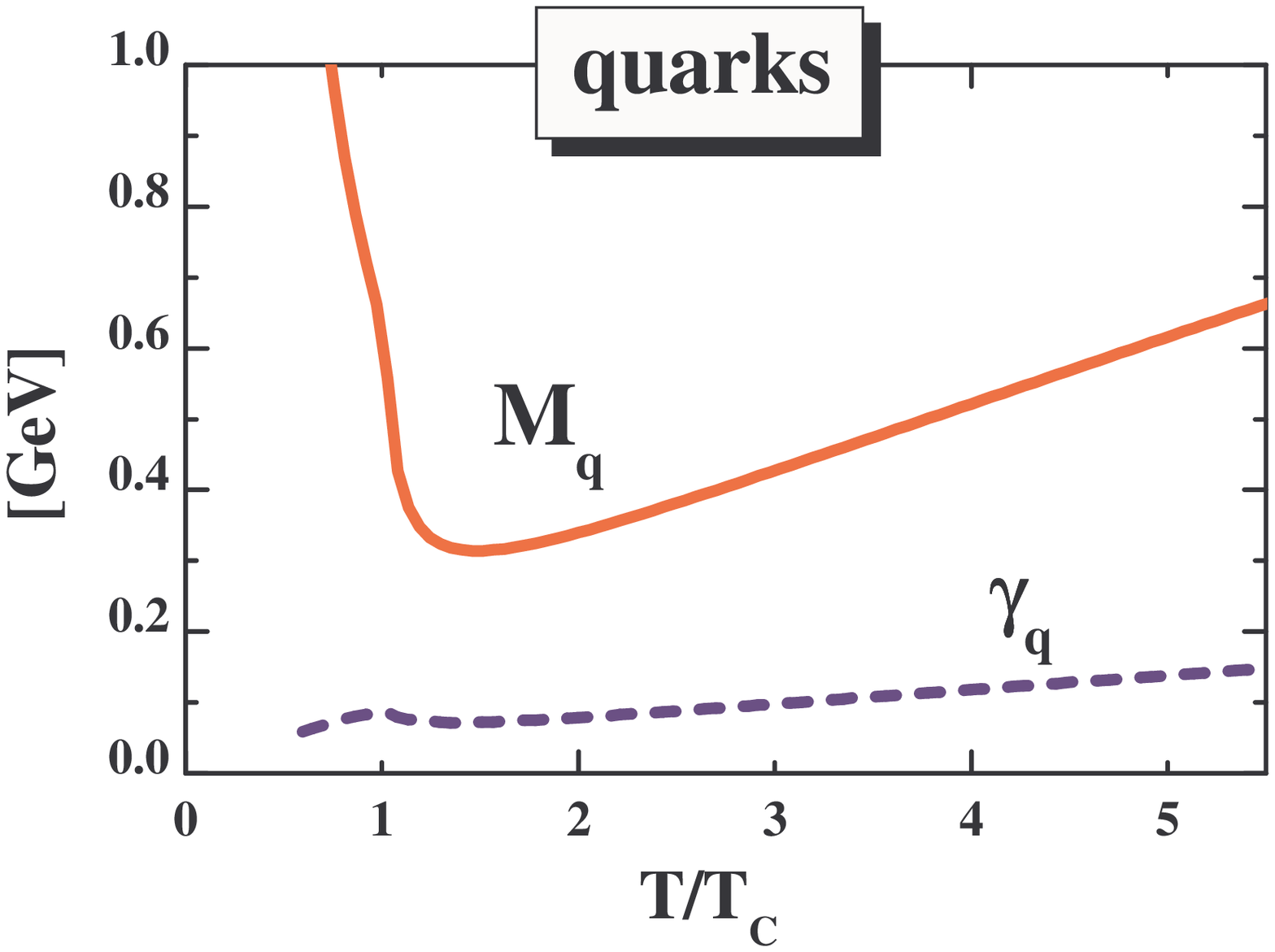}
\caption{ The effective gluon mass
$M_g$ and witdh $\gamma_g$ as function of temperature $T/T_c$ (l.h.s.).
The r.h.s. shows the corresponding quantities for quarks.} \label{fig1}
\end{figure}

The actual gluon mass $M_g$ and width $\gamma_g$ -- employed in
the further calculations -- as well as the quark mass $M_q$ and width $\gamma_q$
are depicted in Fig. \ref{fig1} on the lhs and rhs, respectively, as a
function of $T/T_c$. These values for the masses and widths are
smaller than those presented in Ref. \cite{Andre05} since the
recent lQCD calculations \cite{Cheng08} have been performed with
much smaller bare fermion masses. This, in fact, leads to an
increase in the entropy density $s(T/T_c)$ relative to earlier
lQCD calculations, which implies that the quasiparticle mass and
width - fitting the lQCD results -  become smaller relative to
Refs. \cite{Cassing06,Cassing07,Andre05}.

In line with Ref. \cite{Andre05} the parton spectral functions are
no longer $\delta-$ functions in the invariant mass squared but
taken as
\begin{eqnarray}
 \rho_j(\omega)
 =
 \frac{\gamma_j}{ E_j} \left(
   \frac{1}{(\omega-E_j)^2+\gamma_j^2} - \frac{1}{(\omega+E_j)^2+\gamma_j^2}
 \right)
 \label{eq:rho}
\end{eqnarray} separately for quarks and gluons ($j=q,\bar{q},g$).
With the convention $E^2(\bm p) = \bm p^2+M_j^2-\gamma_j^2$, the
parameters $M_j^2$ and $\gamma_j$ are directly related to the real
and imaginary parts of the  retarded self-energy, e.g. $\Pi_j =
M_j^2-2i\gamma_j\omega$. The spectral function (\ref{eq:rho}) is
antisymmetric in $\omega$ and normalized as \begin{equation}
\label{normalize}
\int_{-\infty}^{\infty} \frac{d \omega}{2 \pi} \ \omega \
\rho_j(\omega, {\bf p}) = \int_0^{\infty} \frac{d \omega}{2 \pi} \ 2 \omega \
\rho_j(\omega, {\bf p}) = 1 \ .
\end{equation}
With the spectral functions fixed by Eqs. (1)-(5) the total energy
density in the DQPM $T^{00}_+ + T^{00}_-$ can be evaluated as
\cite{Cassing07}
\begin{equation} \label{ener}
T^{00}_{\pm} = d_g \int_0^\infty  \frac{d\omega}{2 \pi}
\int \frac{d^3 p}{(2 \pi)^3}\ 2 \omega^2 \rho_g(\omega, {\bf p})
\ n_B(\omega/T) \ \Theta(\pm p^2) \end{equation} $$+ d_q \int_0^\infty  \frac{d\omega}{2 \pi}
\int \frac{d^3 p}{(2 \pi)^3} \ 2 \omega^2 \rho_q(\omega, {\bf p})
\ n_F((\omega-\mu_q)/T) \ \Theta(\pm p^2)$$
 $$+ d_{\bar q} \int_0^\infty  \frac{d\omega}{2 \pi}
\int \frac{d^3 p}{(2 \pi)^3} \ 2 \omega^2 \rho_{\bar q}(\omega, {\bf p})
\ n_F((\omega+\mu_q)/T) \ \Theta(\pm p^2) \ ,$$
where $n_B$ and $n_F$ denote the Bose and Fermi
functions, respectively, while $\mu_q$ stands for the quark chemical potential.
The number of transverse gluonic degrees
of freedom is $d_g=16$ while the fermion degrees of freedom amount
to $d_q=d_{\bar q}=2 N_c N_f=18$ in case of three flavors
($N_f$=3). The indices $\pm$ stand for the time-like (+) and
space-like part ($-$) of the energy density (\ref{ener}) as defined
via $\Theta(\pm p^2)$ with $p^2 = \omega^2 - {\bf p}^2$.

The pressure $P(T)$ then can be obtained by integrating the
differential thermodynamic relation (for $\mu_q=0$) \begin{equation} P -
T \frac{\partial P}{\partial T}= - T^{00} =  P - Ts \end{equation} with the
entropy density $s(T)$ given by
\begin{equation}
s = \frac{\partial P}{\partial T} = \frac{T^{00}+P}{T} \ .
\end{equation} This approach is thermodynamically consistent and
represents a two-particle irreducible (2PI) approximation to hot QCD.

For the further presentation of the DQPM  it is useful to
introduce the shorthand notations (following Ref. \cite{Cassing07})
\begin{equation} \label{conv} \hspace{1.5cm}
 {\rm \tilde Tr}^{\pm}_g \cdots
 =
 d_g\!\int\!\!\frac{d \omega}{2 \pi} \frac{d^3p}{(2 \pi)^3}\,
 2\omega\, \rho_g(\omega)\, \Theta(\omega) \, n_B(\omega/T) \ \Theta(\pm p^2) \, \cdots
 \,\end{equation}
$$   {\rm \tilde Tr}^{\pm}_q \cdots
 =
 d_q\!\int\!\!\frac{d \omega}{2 \pi} \frac{d^3p}{(2 \pi)^3}\,
 2\omega\, \rho_q(\omega)\, \Theta(\omega) \, n_F((\omega-\mu_q)/T) \ \Theta(\pm p^2) \, \cdots
 \,$$ $$  {\rm \tilde Tr}^{\pm}_{\bar q} \cdots =
 d_{\bar q}\!\int\!\!\frac{d \omega}{2 \pi} \frac{d^3p}{(2 \pi)^3}\,
 2\omega\, \rho_{\bar q}(\omega)\, \Theta(\omega) \, n_F((\omega+\mu_q)/T) \ \Theta(\pm p^2) \, \cdots
$$

\noindent
 with $p^2= \omega^2-{\bf p}^2$ denoting the invariant mass
squared. Here the $\Theta(\pm p^2)$-functions project on
'time-like' (+) and 'space-like' ($-$) sectors of the four-momentum.
The traces in (\ref{conv}) then give directly time-like and
space-like 'densities' for gluons $N_g^\pm$ as well as for quarks
$N_q^\pm$ (and antiquarks). Note, that only the time-like parts
have the physical interpretation of particles per volume.
We mention that scalar gluon and
quark (+antiquark) densities are evaluated as
\begin{equation} \label{scalpart}
 \rho_g^s(\frac{T}{T_c})= Tr_g^+ \left( \frac{\sqrt{p^2}}{\omega} \right), \hspace{2cm}
 \rho_q^s(\frac{T}{T_c})= Tr_q^+ \left(\frac{\sqrt{p^2}}{\omega} \right)
+Tr_{\bar q}^+ \left(\frac{\sqrt{p^2}}{\omega} \right) \ .
 \end{equation}
As seen from Fig. 2 in Ref. \cite{CaBra09}  the 'densities' roughly scale with
$(T/T_c)^3$ except for the region close to $T_c$.

\begin{figure}[t]
\centering \includegraphics*[width=85mm]{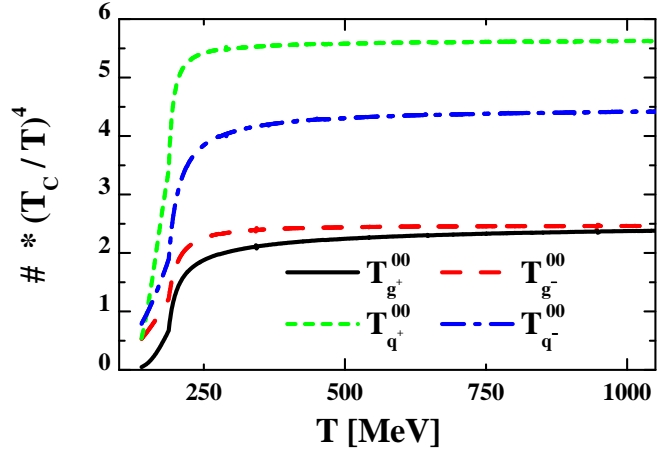} \caption{The
different time- and space-like contributions to the total energy
density  as a function of $T/T_c$ in units of GeV/fm$^3$.
All quantities here have been multiplied by $(T_c/T)^4$ to map out
the scaling properties with temperature $T$.} \label{fig3}
\end{figure}

The time- and space-like energy densities $T^{00}_{\pm}$ for
gluons and quarks (+ antiquarks) are displayed in Fig. \ref{fig3}
as a function of $T/T_c$ for the present parameter set in units of GeV/fm$^3$.
The space-like energy density for gluons
is slightly larger than the time-like part whereas the fermion contribution (quarks +
antiquarks) is clearly dominated by the time-like sector. All
quantities roughly scale as $(T/T_c)^4$ except for the region
close to $T_c$.

\begin{figure}[t]
\vspace{0.5cm}
 \includegraphics*[width=73mm]{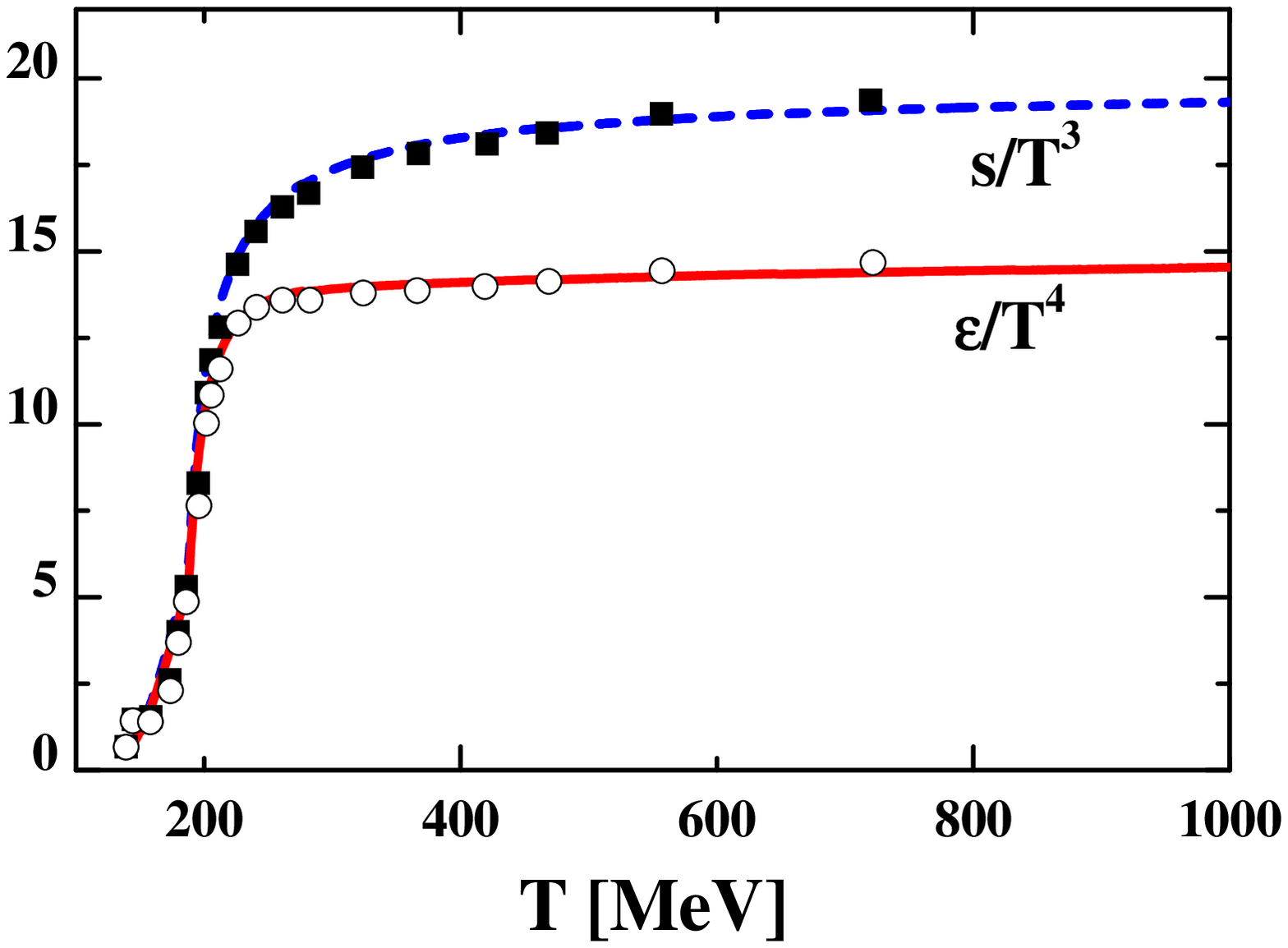} \hspace{0.3cm}
 \includegraphics*[width=80mm]{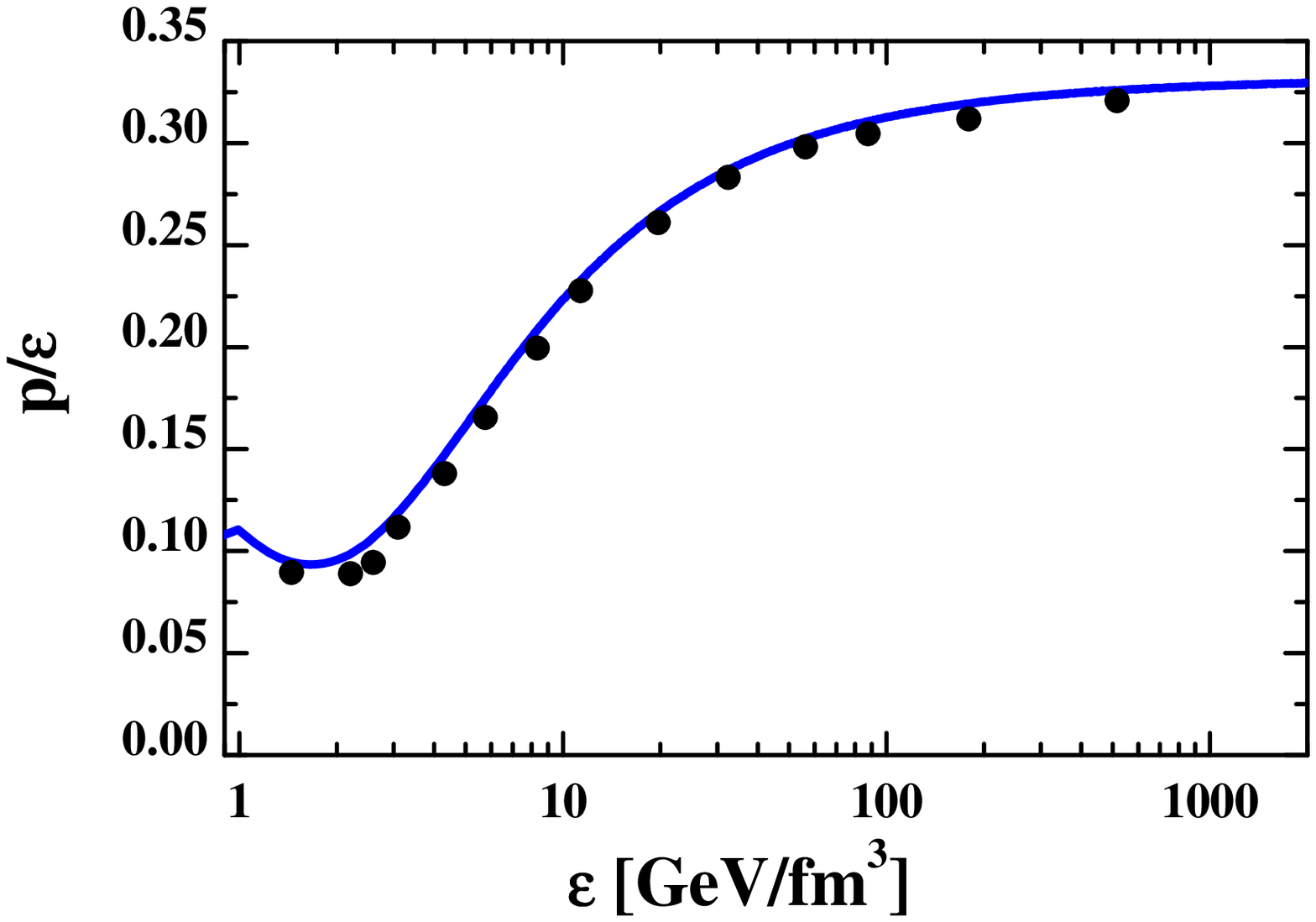}
 \caption{Lhs:
The entropy density $s(T)$ (dashed line) and energy
density $\epsilon(T)$ (solid line) from the DQPM in comparison to
the lQCD results from Ref. \cite{Cheng08} (full squares and open dots).
Rhs: The
equation-of-state (solid line) from the DQPM in comparison to the
lQCD results from Ref. \cite{Cheng08} (full dots).} \label{fig4}
\end{figure}

A direct comparison of the entropy density $s(T)$ and energy
density $\epsilon(T)$ from the DQPM with results from lQCD from
Ref. \cite{Cheng08} is presented in Fig. \ref{fig4} (lhs). Both results
have been divided by $T^3$ and $T^4$, respectively, to
demonstrate the scaling with temperature. We briefly note that the
agreement is sufficiently good. This also holds for the
'equation of state', i.e. $P/\epsilon$ versus $\epsilon$ as
demonstrated in Fig. \ref{fig4} (rhs).

As discussed in detail in Refs. \cite{Cassing06,Cassing07} and
explicitly addressed in Eq. (\ref{ener}), the energy-density
functional can be separated in space-like and time-like sectors
when the spectral functions acquire a finite width. The space-like
(-) part of (\ref{ener})
\begin{equation} \label{Vp}
V_p(T,\mu_q) = T^{00}_{g-} + T^{00}_{q-} + T^{00}_{{\bar q}-} \end{equation}
may be interpreted as
a potential-energy density $V_p$ since the field quanta involved are
virtual and correspond to partons exchanged in interaction
diagrams. The time-like part (+) of (\ref{ener}) corresponds to effective field quanta
which can be propagated within the light-cone. Without repeating the
details we mention that mean-field potentials for partons can be
defined by the derivative of the potential-energy density $V_p$ with
respect to the time-like parton densities and effective interactions
by second derivatives of $V_p$ (cf. Section 3 in Ref. \cite{Cassing07}).

\begin{figure}[t]
\includegraphics*[width=75mm]{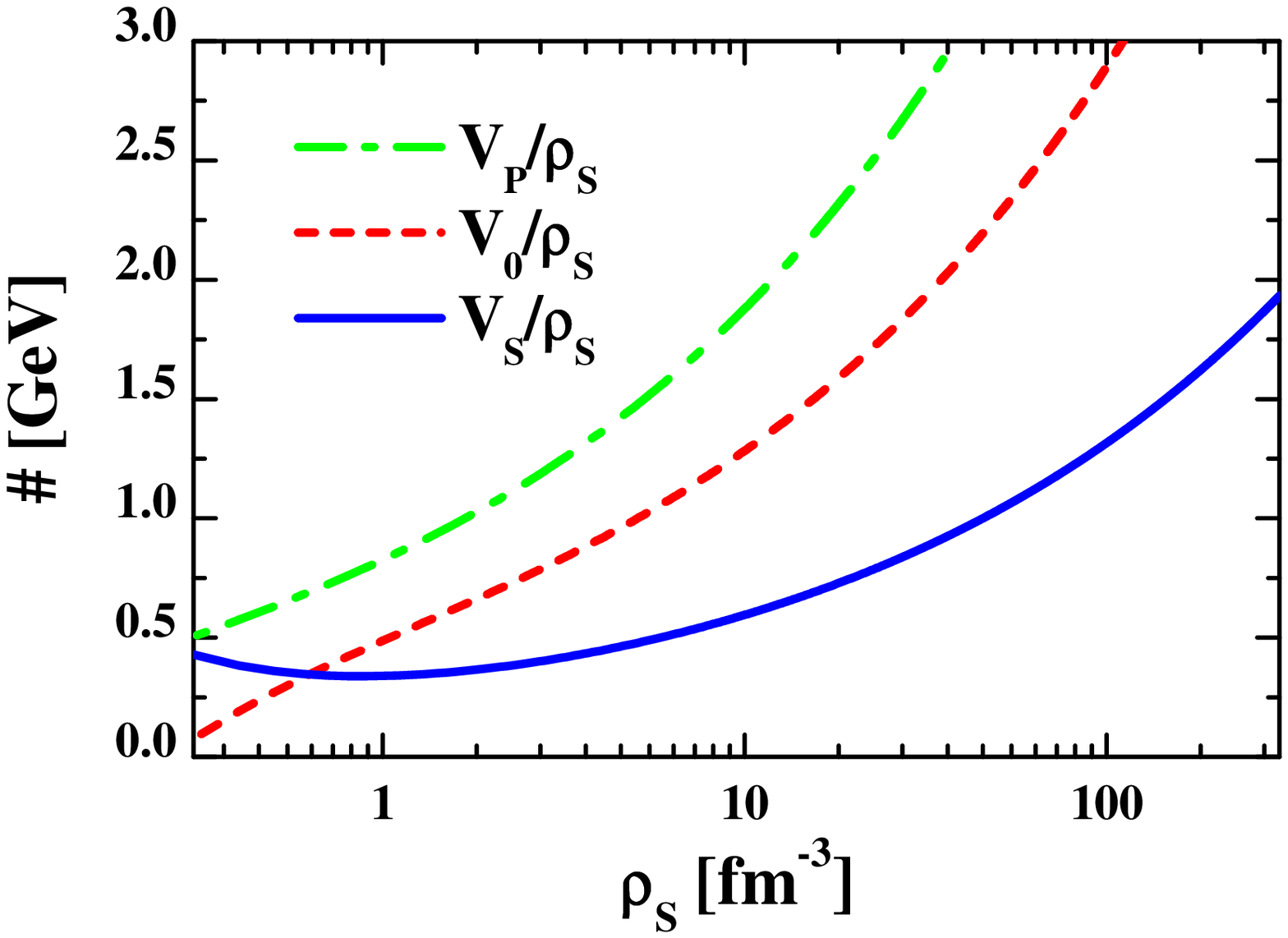} \hspace{0.3cm}
\includegraphics*[width=75mm]{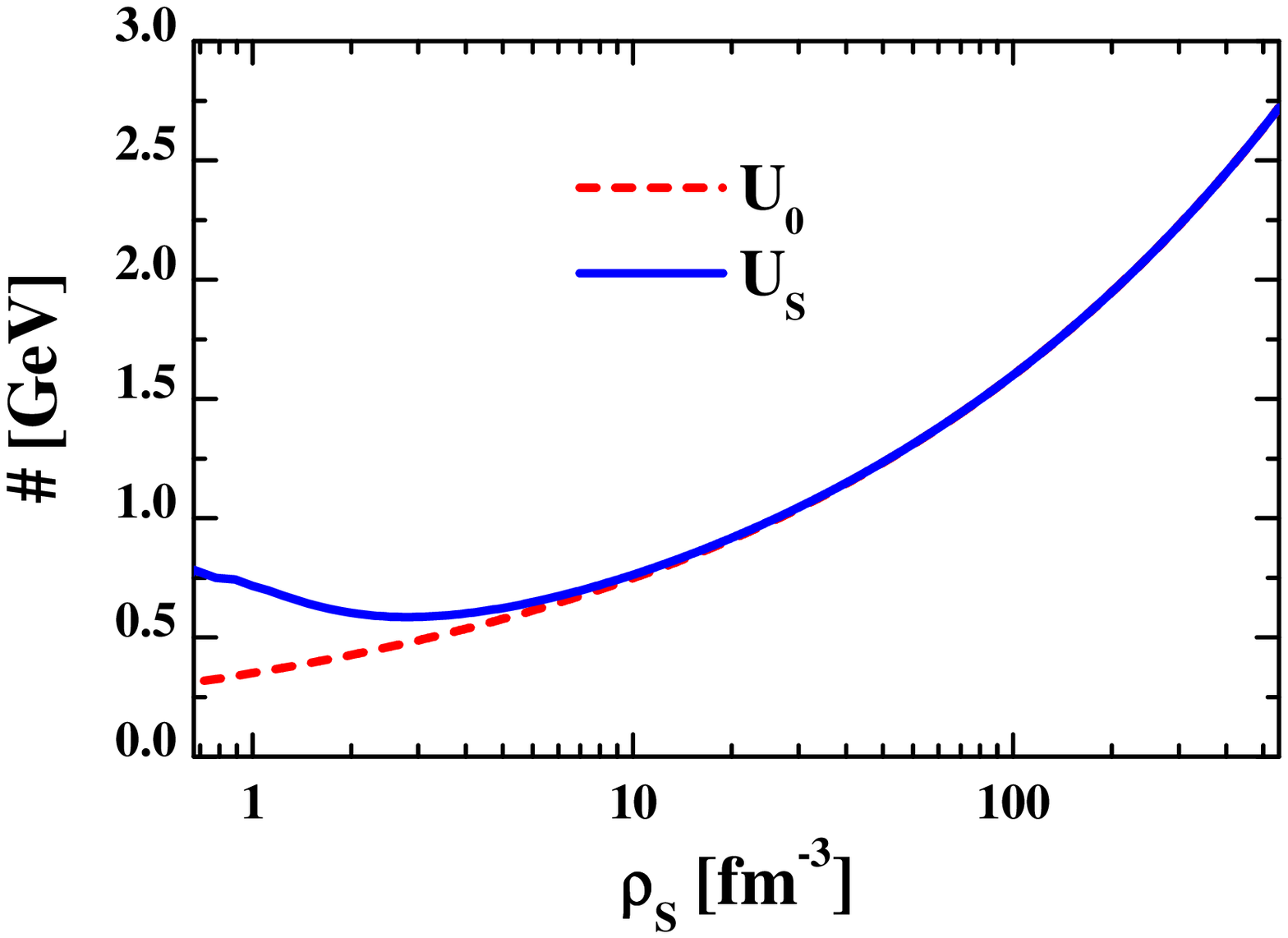}
\caption{Lhs: The total potential energy per scalar parton (green dashed-dotted line)
separated into scalar (solid blue line) and vector (red dashed line)
parts (see text) as a function of the scalar parton density
$\rho_s$. Rhs:   The  scalar (blue solid line) and vector (red dashed line)
mean-fields for quarks  (\ref{MF}) as a function of the scalar
parton density $\rho_s$.} \label{fig6}
\end{figure}

In analogy to relativistic effective approaches for nucleonic
degrees of freedom \cite{Walecka,Maruyama} we assume the potential energy
density (\ref{Vp}) to be a sum of  scalar and vector parts, i.e.
\begin{equation} \label{Vp2}
V_p(T,\mu_q) = V_s(T,\mu_q) + V_v(T,\mu_q) \ . \end{equation} In
the dynamical quasiparticle picture the pressure $P$ then is a
sum of kinetic as well as  interaction parts, i.e.
\begin{equation} \label{pres2} P(T,\mu_q) = \langle P_{xx} \rangle_{T,\mu_q} -
V_s(T,\mu_q) + V_v(T,\mu_q) \ \end{equation} with
\begin{equation} \label{k1}\langle P_{xx} \rangle_{T,\mu_q} = \frac{1}{3} \left(Tr_g^+ \left( \frac{{\bf
p}^2}{\omega} \right) + Tr_q^+ \left( \frac{{\bf p}^2}{\omega}
\right) + Tr_{\bar q}^+ \left( \frac{{\bf p}^2}{\omega} \right)
\right) \ .
\end{equation}
In a similar way the total energy density $\epsilon$ can be
expressed as
\begin{equation} \label{k2}
\epsilon(T,\mu_q) = \langle \omega \rangle_{T, \mu_q} + V_s(T,\mu_q) +
V_v(T,\mu_q) \end{equation} with the time-like quasiparticle
energy density given by
\begin{equation} \label{p2} \langle \omega \rangle_{T,\mu_q} =  Tr_g^+ \left( \omega \right)
+ Tr_q^+ \left( \omega \right) + Tr_{\bar q}^+ \left( \omega
\right) \ .
\end{equation} Note the different signs for the scalar interaction part
in (\ref{pres2}) and (\ref{k2}) which stem from the metric tensor
in the energy-momentum tensor $T^{\mu \nu}$. Since the total energy
density as well as pressure are known in the DQPM and the kinetic
parts (\ref{k1}) and (\ref{k2}) can be evaluated as well, the
vector and scalar interaction densities can be uniquely extracted
from the equations above. The corresponding results are displayed
in Fig. \ref{fig6} (lhs) for the total interaction per scalar particle
$V_p/\rho_s$ (upper dot-dashed line) and its decomposition in
scalar (solid line) and vector (dashed line) parts as a function
of the scalar parton density $\rho_s$,
\begin{equation}
\rho_s= \rho^s_g + \rho^s_q + \rho^s_{\bar q} ,
\end{equation}
which is a convenient (Lorentz invariant) quantity to characterize the
system instead of the Lagrange parameters $T^{-1}$ and $\mu_q$. In
fact, as found in Ref. \cite{Crev} the thermodynamic quantities for different $T$ and
$\mu_q$ are very close when representing them as a function of
$\rho_s$ or the parton density $\rho_p$, respectively.

Scalar and vector mean-field potentials are defined by derivatives
of $V_s$ and $V_v$ with respect to the densities $\rho_s$ and
$\rho_v$, respectively,
\begin{equation} \label{MF}
U_s(\rho_s) = \frac{\partial V_s}{\partial \rho_s} ; \hspace{1cm}
U_v^0 = \frac{\partial V_v}{\partial \rho_p} \ .
\end{equation}
Both quantities (after differentiation)  may be considered as a function of the
parton-scalar density $\rho_s$. The resulting functions are displayed in
Fig. \ref{fig6} (rhs) as a function of $\rho_s$ and
show approximately equal results at high scalar density (or energy
density, respectively). The low-density properties, however, are
very different. Whereas the vector-mean field drops to zero with
decreasing $\rho_s$ the scalar-mean field increases substantially
below $\rho_s \approx$ 1 fm$^{-3}$; it very smoothly increases
with density for $\rho_s > $ 2 fm$^{-3}$ \footnote{Note the logarithmic scale in
$\rho_s$.}. Accordingly scalar forces on the partons $\sim - \partial
U_s/\partial \rho_s \nabla \rho_s$ are rather small in the high-density
partonic phase and only become stronger in the low-density phase
close to hadronization. The DQPM as well as lQCD do not allow to
extrapolate $U_s$ down to $\rho_s = 0$ reliably  but one might imagine that
$U_s(\rho_s \rightarrow 0) \rightarrow \infty$ thus encoding
scalar confinement on the mean-field level. We note that in actual
PHSD calculations such low partonic densities are not probed
dynamically because the partons hadronize in the region $\rho_s
\approx$ 1 to 2 fm$^{-3}$ (see below).

With the scalar- and vector-mean fields fixed for the partons
(assuming the gluon mean fields to be twice the quark mean fields
according to the analysis within the DQPM (cf. Ref. \cite{Cassing07})
and with the effective masses and dynamical widths fixed by the
DQPM (as a function of $\rho_s$ instead of $(T/T_c)$) the
mean-field propagation of partons in PHSD is fully determined by
the off-shell transport equations (cf. the review \cite{Crev}).
Note that these mean-fields are presently momentum
independent whereas nuclear physics studies suggest a moderate
dependence on the particle momentum \cite{Lari,Rappi}.

The transport equations still require to specify
the elastic and inelastic cross sections of quarks and gluons,
which enter the collision terms. For the latter quantities we
refer the reader to Ref. \cite{CaBra09}.

\subsection{Hadronization in PHSD}
 The hadronisation,
i.e. the transition from partonic to hadronic degrees of freedom,
is described in PHSD by local covariant transition rates as
introduced in Ref.
\cite{PRC08} e.g. for $q+\bar{q}$ fusion to a meson $m$ of
four-momentum $p= (\omega, {\bf p})$ at space-time point
$x=(t,{\bf x})$:
\begin{eqnarray}
&&\phantom{a}\hspace*{-5mm} \frac{d N_m(x,p)}{d^4x d^4p}= Tr_q
Tr_{\bar q} \
  \delta^4(p-p_q-p_{\bar q}) \
  \delta^4\left(\frac{x_q+x_{\bar q}}{2}-x\right) \nonumber\\
&& \times \omega_q \ \rho_{q}(p_q)
   \  \omega_{\bar q} \ \rho_{{\bar q}}(p_{\bar q})
   \ |v_{q\bar{q}}|^2 \ W_m(x_q-x_{\bar q},(p_q-p_{\bar q})/2) \nonumber \\
&& \times N_q(x_q, p_q) \
  N_{\bar q}(x_{\bar q},p_{\bar q}) \ \delta({\rm flavor},\, {\rm color}).
\label{trans}
\end{eqnarray}
In Eq. (\ref{trans}) we have introduced the shorthand notation,
\begin{equation}
Tr_j = \sum_j \int d^4x_j \int \frac{d^4p_j}{(2\pi)^4} \ ,
\end{equation}
where $\sum_j$ denotes a summation over discrete quantum numbers
(spin, flavor, color); $N_j(x,p)$ is the phase-space density of
parton $j$ at space-time position $x$ and four-momentum $p$.  In
Eq. (\ref{trans}) $\delta({\rm flavor},\, {\rm color})$ stands
symbolically for the conservation of flavor quantum numbers as
well as color neutrality of the formed hadron $m$ which can be
viewed as a color-dipole or 'pre-hadron'.  Furthermore, $v_{q{\bar
q}}(\rho_p)$ is the effective quark-antiquark interaction  from
the DQPM  (displayed in Fig. 10 of Ref. \cite{Cassing07}) as a
function of the local parton ($q + \bar{q} +g$) density $\rho_p$
(or energy density). Furthermore, $W_m(x,p)$ is the dimensionless phase-space
distribution of the formed 'pre-hadron', i.e.
\begin{equation} \label{Dover} W_m(\xi,p_\xi) =
\exp\left( \frac{\xi^2}{2 b^2} \right)\ \exp\left( 2 b^2 (p_\xi^2- (M_q-M_{\bar
q})^2/4) \right)
\end{equation} with $\xi = x_1-x_2 = x_q - x_{\bar q}$ and $p_\xi = (p_1-p_2)/2
= (p_q - p_{\bar q})/2$ (which has been previously introduced in
Eq. (2.14) of Ref. \cite{Dover}). The width parameter $b$ is fixed
by $\sqrt{\langle r^2 \rangle} = b$ = 0.66 fm (in the rest frame) which
corresponds to an average rms radius of mesons. We note that the
expression (\ref{Dover}) corresponds to the limit of independent
harmonic oscillator states and that the final hadron-formation
rates are approximately independent of the parameter $b$ within
reasonable variations. By construction the quantity (\ref{Dover})
is Lorentz invariant; in the limit of instantaneous 'hadron
formation', i.e. $\xi^0=0$, it provides a Gaussian dropping in the
relative distance squared $({\bf r}_1 - {\bf r}_2)^2$. The
four-momentum dependence reads explicitly (except for a factor
$1/2$)
\begin{equation} (E_1 - E_2)^2 - ({\bf p}_1 - {\bf p}_2)^2 -
(M_1-M_2)^2 \leq 0
\end{equation} and leads to a negative argument of the second
exponential in (\ref{Dover}) favoring the fusion of partons with
low relative momenta $p_q - p_{\bar q}= p_1-p_2$.

Related transition rates (to Eq. (\ref{trans})) are defined for
the fusion of three off-shell quarks ($q_1+q_2+q_3 \leftrightarrow
B$) to color neutral baryonic ($B$ or $\bar{B}$) resonances of
finite width (or strings) fulfilling energy and momentum
conservation as well as flavor current conservation using Jacobi coordinates
(cf. Ref. \cite{CaBra09}).

On the hadronic side the PHSD transport approach includes explicitly the  baryon octet
and decouplet, the $0^-$- and $1^-$-meson nonets as well as
selected higher resonances as in HSD \cite{HSD}. Hadrons of higher
masses ($>$ 1.5 GeV in case of baryons and $>$ 1.3 GeV in case of
mesons) are treated as 'strings' (color-dipoles) that  decay to
the known (low-mass) hadrons
according to the JETSET algorithm \cite{JETSET}. We discard an
explicit recapitulation of the string decay and refer the reader
to the original work \cite{JETSET} or Ref. \cite{Falter}.

\section{Application to nucleus-nucleus collisions}
In this Section we employ the PHSD approach - described in Section
2 - to nucleus-nucleus collisions at moderate
relativistic energies, i.e. at SPS energies where our approximations are
expected to work. Note that at RHIC or LHC energies
other initial conditions (e.g. a color-glass condensate \cite{Larry}) might
be necessary. Since this is a slightly different subject we here restrict to
bombarding energies below 160 A$\cdot$GeV where such problems/questions are expected to
be not relevant.

\begin{figure}[thb]
\includegraphics*[width=75mm]{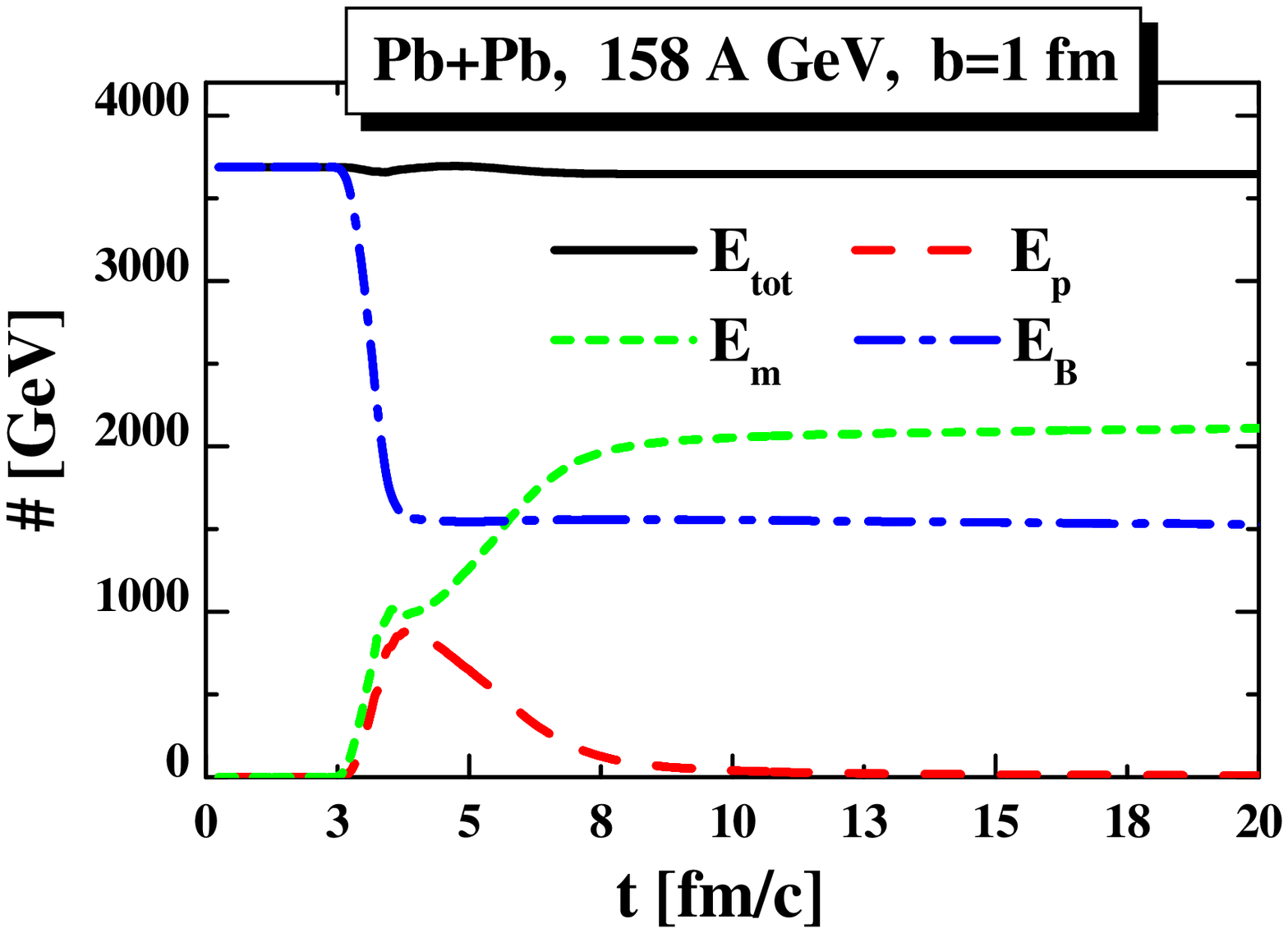} \hspace{0.5cm}
\includegraphics*[width=75mm]{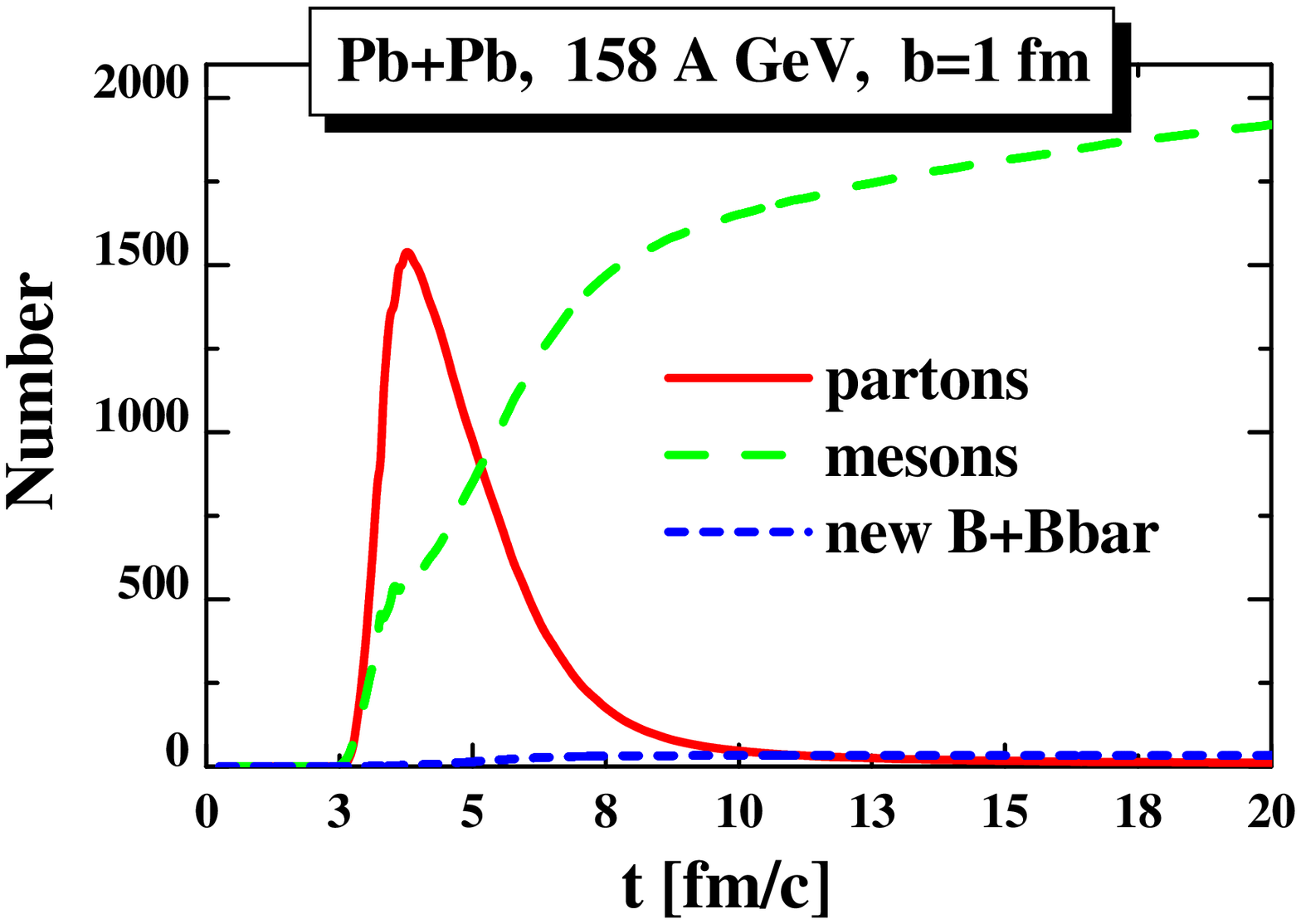}
\caption{Lhs: The total energy $E_{tot}$
(upper solid line) for central ($b$=1 fm) collisions of Pb+Pb at 158 A$\cdot$GeV.
The long-dashed (red) line shows the energy contributions $E_p$ from
partons while the short- dashed (green) line displays the energy
contribution from mesons $E_m$ (including 'unformed mesons' in strings).
The dot-dashed (blue) line is the contribution of
baryons (and antibaryons) $E_B$; the difference between the initial
baryonic energy $E_B(t=0)$ and final baryonic energy $E_B(t
\rightarrow \infty$) gives the energy that is converted during the
heavy-ion collision to final mesonic states. Rhs: The number
of produced partons (solid red line), mesons (long dashed green
line) and newly produced baryon + antibaryons (blue dashed line)
as a function of time  for Pb+Pb at 158 A$\cdot$GeV (for $b$=1
fm). Note that the number of mesons still increases for $t > $ 20
fm/c due to the decay of vector mesons.} \label{fig8}
\end{figure}

\subsection{Partonic energy fractions}
We start with a consideration of energy
partitions in order to map out the fraction of partonic energy in
time for relativistic nucleus-nucleus collisions.
In Fig. \ref{fig8} (lhs) we show the energy balance for a central
(impact parameter $b$=1 fm) reaction of Pb+Pb at 158 A $\cdot$
GeV, i.e. at the top SPS energy. The total energy $E_{tot}$ (upper
line) - which at $t=0$ is given by the energy of the colliding
nuclei in the cms  - is conserved  throughout the
reaction, i.e. in the partonic and hadronization phase as well as
in the hadronic phase.   Whereas in the first $\sim$ 3 fm/c the total energy is
entirely contained in the impinging  nucleons (including about 6
MeV per nucleon of binding energy) a rapid transition to partonic
degrees of freedom is seen at $t \approx 3$ fm/c, i.e. when the
nuclei have started to overlap and react. We recall that the
transition time of the two Pb-nuclei is about $2
R_{Pb}/\gamma_{cm} \approx$ 1.5 fm/c at the top SPS energy. During
this time period about 60\% of initial kinetic energy of nucleons is converted
to partons (red dashed line) and  to mesons (short dashed
green line) in the surface region ('corona') of the colliding
system. Note that in the 'mesonic' energy  $E_m$ also 'unformed
mesons' - as fragments of the strings - are accounted for. The
energy of residual baryons (including antibaryons) is shown in
terms of the dot-dashed (blue) line and is almost constant for $t
> $ 5 fm/c implying that the various final-state interactions do
not show up significantly in the energy fractions. The partonic
phase - in a limited space-time region - approximately ends for $t
>$ 9 fm/c which means that the further time evolution of the
system is essentially described by hadronic interactions (HSD).
Note that a sizeable fraction of energy is asymptotically still
contained in the baryons. Since the baryon-rest masses amount to
an energy of about 449 GeV (including newly produced $B \bar{B}$
pairs) this implies that full stopping is not achieved in central
Pb+Pb collisions at 158 A$\cdot$ GeV.

Let's have a closer look at the 'particle' composition in time for
this reaction. We concentrate on those species that carry the energy
transferred during the collision to new degrees of freedom.
In this respect we display in Fig. \ref{fig8} (rhs) the number of
produced partons (solid red line), mesons (long dashed green line) and
newly produced baryons + antibaryons (blue dashed line) as a
function of time for the same reaction as before. We recall that
the initial number of nucleons is 416 in this case. Slightly more
than 1500 partons are produced during the passage time of the
nuclei which disappear practically after 9 fm/c and essentially form mesons.
The number of newly produced $B +\bar{B}$ pairs is small at this
energy but its flavor decomposition is quite interesting.
An essential point here is that the number of final
hadronic states is larger than the number of partons, i.e. there
is a production of entropy in the hadronization process as pointed
out before in Ref. \cite{PRC08}. This implies that in
PHSD the second law of thermodynamics is not violated in the
hadronization process!

\subsection{Strangeness in PHSD}
As found in Ref. \cite{CaBra09} the impact of the partonic degrees
of freedom in PHSD on the longitudinal rapidity distribution of protons,
pions and kaons is only small in central Pb+Pb collisions at SPS energies from 40 to 158 A$\cdot$GeV.

\begin{figure}[tb]
\centerline{\includegraphics*[width=125mm]{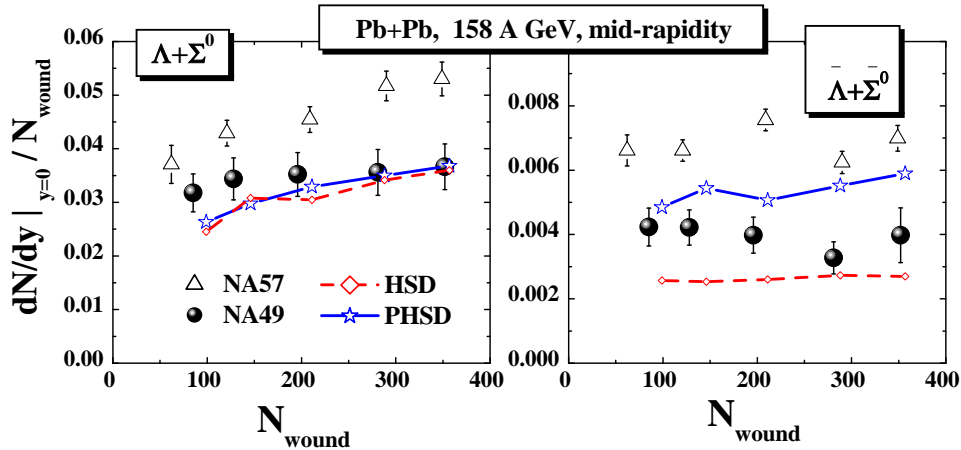}}
\caption{The multiplicities of $(\Lambda + \Sigma^0)/N_{wound}$ (l.h.s.) and
$(\bar \Lambda + \bar \Sigma^0)/N_{wound}$ (r.h.s.) as a function of the number of wounded
nucleons $N_{wound}$ for Pb+Pb collisions at 158 A$\cdot$GeV at mid-rapidity from PHSD
(blue solid lines) and HSD (red dashed-dotted lines) in comparison to the
experimental data from the NA57 Collaboration \cite{NA57} (open triangles)
and the NA49 Collaboration \cite{NA49_aL09} (solid dots).  The HSD and PHSD calculations
have an error of about 5$-$10\% due to limited statistics.  }
\label{fig15d}
\end{figure}

Additional experimental information is provided by the centrality
dependence of the strange (and antistrange) baryon yield. In this
respect we compare in Fig. \ref{fig15d} the multiplicities of $(\Lambda
+ \Sigma^0)/N_{wound}$ (l.h.s.) and $(\bar \Lambda + \bar
\Sigma^0)/N_{wound}$ (r.h.s.) as a function of the number of wounded
nucleons $N_{wound}$ for Pb+Pb collisions at 158 A$\cdot$GeV at
mid-rapidity from PHSD (blue solid lines) and HSD (red dashed-dotted
lines) to the experimental data from the NA57 Collaboration \cite{NA57}
(open triangles) and the NA49 Collaboration \cite{NA49_aL09} (solid
dots).  The (green) full squares correspond to the 10\% central data
points at midrapidity from Ref. \cite{NA49b}. We mention that we employ
the same definition of wounded nucleons $N_{wound}$ as the NA49
Collaboration.  Whereas the HSD and PHSD calculations both give a
reasonable description of the $\Lambda + \Sigma^0$ yield of the NA49
Collaboration, both models underestimate the NA57 data (open triangles)
by about 30\%. An even larger discrepancy in the data from the NA49 and
NA57 Collaborations is seen for $(\bar \Lambda + \bar \Sigma^0)/N_{wound}$
(r.h.s.); here the PHSD calculations give results which are in between the
NA49 data (solid dots) and the NA57 data (open triangles). We also see that HSD
underestimates the $(\bar \Lambda + \bar \Sigma^0)$ midrapidity yield
at all centralities.

The latter results suggest that the partonic phase does not show up
explicitly in an enhanced production of strangeness (or in particular
strange mesons and baryons) but leads to a different redistribution of
antistrange quarks between mesons and antibaryons in the hadronization process.

\section{Summary}

In this contribution we have presented a brief overview of the DQPM model
and its results in comparison to lattice QCD. Furthermore, we have addressed
relativistic collisions of Pb+Pb at SPS energies in the PHSD approach which includes explicit
partonic degrees of freedom as well as dynamical local transition
rates from partons to hadrons (\ref{trans}). The
hadronization process conserves four-momentum and all flavor
currents and slightly increases the total entropy  since the
'fusion' of rather massive partons dominantly leads to the
formation of color neutral strings or resonances that decay
microcanonically to lower mass hadrons. Since this dynamical
hadronization process increases the total entropy the
second law of thermodynamics is not violated (as is the case for simple
coalescence models incorporating massless partons).

The PHSD approach has been also applied to nucleus-nucleus collisions
from 40 to 160 A$\cdot$GeV in order to explore the space-time
regions of 'partonic matter' \cite{CaBra09}. We have found that even central
collisions at the top SPS energy of $\sim$158 A$\cdot$ GeV show a
large fraction of non-partonic, i.e. hadronic or string-like
matter, which can be viewed as a 'hadronic corona'.
This finding implies that neither purely hadronic
nor purely partonic 'models' can be employed to extract physical
conclusions in comparing model results with data.

On the other
hand - studying in detail Pb+Pb reactions at 40, 80 and 158
GeV$\cdot$GeV in comparison to the data from the NA49
Collaboration \cite{CaBra09}- it is found that the partonic phase has only a
very low impact on rapidity distributions of hadrons but a
sizeable influence on the transverse-mass distribution of final
kaons due to the parton interactions.
The most pronounced effect is seen  on the production of multi-strange
antibaryons due to a slightly enhanced $s{\bar s}$ pair production in
the partonic phase from massive time-like gluon decay and a more
abundant formation of strange antibaryons in the hadronization process. This
enhanced formation of strange antibaryons in central Pb+Pb collisions
at SPS energies by hadronization supports the early suggestion by
Braun-Munzinger and Stachel \cite{PBM,PBM2} in  the statistical
hadronization model - which describes well particle ratios from AGS to
RHIC energies.

Some note of caution has to be stated here with respect to
applications of PHSD at FAIR energies (5-40 A$\cdot$GeV) since the partonic
equation-of-state employed so far - as fixed to the lQCD results from Ref.
\cite{Cheng08} - describes a crossover transition between the
hadronic and partonic phase while at lower SPS and FAIR energies a
first-order phase transition and the appearance of a critical
point in the QCD phase diagram are expected \cite{CBMbook}. Such
phenomena may not be described by the present realization of PHSD
but need subtle extensions. Additionally, we have to
mention that the high transition temperature from Ref.
\cite{Cheng08} might be questioned since related lattice
calculations of the Wuppertal group in 2009 lead to lower critical
temperatures \cite{Wupper}. Once the question of the 'proper $T_c$
or critical energy density $\epsilon_c$ is settled by the lattice
community we might have
to readjust the partonic equation-of-state in PHSD accordingly.
On the other hand an application to
RHIC energies is rather straight forward - except for a possible
color-glass initial state - and detailed results from PHSD will be
presented in a forthcoming study.

\section*{Acknowledgement}
The authors are grateful to C. Blume, C. H\"ohne, O. Linnyk and S. Mattiello for valuable
discussions.


\end{document}